\documentclass[12pt]{article}
\pdfoutput=1

\usepackage{amssymb,amsmath,mathrsfs,epstopdf,slashed,color}
\usepackage{tikz}
\usetikzlibrary{matrix}
\usetikzlibrary{positioning}
\usepackage{fancyhdr} 
\usepackage{lastpage} 
\usepackage{extramarks} 
\usepackage{graphicx} 
\usepackage{lipsum} 
\usepackage{amsmath}
\usepackage{amsfonts}
\usepackage{amssymb}
\usepackage{latexsym}
\usepackage{color}
\usepackage{xypic}
\usepackage{cancel,slashed}
\usepackage[linktocpage]{hyperref}
\usepackage{booktabs}
\usepackage{caption}
\usepackage{comment}
\usepackage{csquotes} 
\usepackage{caption}
\usepackage{subcaption}
\usepackage{url}

\usepackage{graphicx}
\usepackage{placeins}
\usepackage{hyperref}
\usepackage{cite}
\usepackage{braket}
\usepackage{bm}
\usepackage{diagbox}
\usepackage{multirow}
\usepackage{subcaption}

\usepackage{xcolor}
\usepackage{xspace}
\allowdisplaybreaks[1]

\makeatletter

\newcommand\blfootnote[1]{%
  \begingroup
  \renewcommand\thefootnote{}\footnote{#1}%
  \addtocounter{footnote}{-1}%
  \endgroup
}
\begin{document}
	
	\begin{titlepage}
		
		\setcounter{page}{0}
		

		
		

		\vskip 1cm
		\begin{center}
			
			
			\vskip 1cm
			
            {\Large WTH! Wok the Hydrogen: \\Measurement of Galactic Neutral Hydrogen in Noisy Urban Environment Using Kitchenware}
			
			\vskip 0.6cm
			 
			{Leo W.H. Fung${}^{1,2}$}, Albert Wai Kit Lau${}^{1,2,3,*}$, {Ka Hung Chan${}^{4}$}, {Ming Tony Shing${}^{1}$}
			
			\vskip 0.6cm

			${}^1$ Department of Physics, \\
			Hong Kong University of Science and Technology, Hong Kong \\
            ${}^2$ Jockey Club Institute for Advanced Study, \\
			Hong Kong University of Science and Technology, Hong Kong \\
			${}^3$ Dunlap Institute for Astronomy \& Astrophysics, University of Toronto, Canada \\
            ${}^4$ Department of Mechanical Engineering, \\
			Hong Kong University of Science and Technology, Hong Kong \\
            \blfootnote{${}^*$ \textit{corresponding author, awklau@connect.ust.hk}} \\
			
			
			\vskip 1.0cm
            \abstract{ \normalsize
            Astronomy observation is difficult in urban environments due to the background noise generated by human activities. Consequently, promoting astronomy in metropolitan areas is challenging. In this work, we propose a low-cost, educational experiment called Wok the Hydrogen (WTH) that offers opportunities for scientific observation in urban environments, specifically the observation of the $21$ cm ($f_{21} = 1420.4$ MHz) emission from neutral hydrogen in the Milky Way. 
            We demonstrate how to construct a radio telescope using kitchenware, along with additional electronic equipment that can be easily purchased online. The total system cost is controlled within \$150. 
            We also outline the subsequent data analysis procedures for deriving the recession velocity of galactic hydrogen from the raw data. 
            The system was tested on the campus of the Hong Kong University of Science and Technology, which is located approximately 2 km northeast of the nearest residential area with a population of 0.4 million and about 10 km east of the downtown area with a population of 2 million. We show that a precision of $\Delta v \approx \pm 20$ km s$^{-1}$ can be achieved for determining the recession velocity of neutral hydrogen with this relatively simple setup, and the precision can be further improved with longer exposure time.
            
            }
			\vspace{5mm}
			\vspace{1cm}
			
		\end{center}
	\end{titlepage}

	\setcounter{page}{1}
	\setcounter{footnote}{0}
	
	\tableofcontents
	
	\parskip=5pt

	\section{Introduction}
    Promotion of astronomy to the public is long known to be challenging, especially in the urban area. 
    The intensive light pollution in the urban environment prevents most observation of astronomical phenomena apart from the sun and the moon.
    Not to mention the observation of the diffuse light from the gas in the Milky Way, which is possible only in remote area, and often requires the help of long exposure time to improve the signal-to-noise ratio (SNR).
    Furthermore, the observation of the diffuse emission from the Milky Way can be assisted by adopting optical filters to selectively filter away the emission bands from human activities. 
    However, filters are often fairly expensive and require extra effort to install in camera systems for astrophotography. 

    Radio observation in a noisy urban area is, however, feasible for the following reasons.
    Owing to the much lower frequency of radio waves compared to optical wavelengths, filtering out unwanted signals can easily be implemented using digital Fourier transform in computer software. This method of separating the signal frequency from the noisy background dramatically increases the SNR.
    A fairly inexpensive hardware filter can also be used as the first stage filter to further prevent unwanted noise from mixing with the desired signal in the baseband.
    Secondly, considering the fairly large beam size, most terrestrial noises are effectively non-directional at large distances. This allows for easy noise subtraction by slightly shifting the pointing of the radio telescope.
    These properties imply that the SNR for radio observation, particularly in narrowband observations, can be drastically improved even in noisy environments. This makes radio observation ideal for practical use in urban areas.

    Apart from concerns about noise figures, radio observation is also ideal in (sub)tropical regions.
    In many (sub)tropical areas (such as Hong Kong where this project is being conducted), the optical visibility is very poor for most of the year due to the climate patterns. The sky is often covered by clouds or experiencing rain, making optical observations of the sky impossible on most days. 
    When organizing astronomical promotion activities, a specific date for observation needs to be fixed in advance. However, there is no guarantee of favorable weather conditions on the proposed date.
    Radio observation is particularly useful in this regard. 
    Thanks to the ability of radio waves (at 21 cm) to penetrate through clouds, astronomical observations are possible almost throughout the year regardless of weather condition. 
    The only exception is during thunderstorms, where the bursty radio signals from lightning may damage the electronics in the radio telescope. 
    The intense energy released from lightning can overdrive the electronic amplifiers, resulting in over-current flowing along the unprotected electronics and damaging the system.
    Another safety concern is that the metallic aperture of the radio telescope can act as a lightning attractor, endangering nearby users of the radio telescope.
    Other than these occasional situations, radio observation can be performed in most weather conditions, making astronomical observations more accessible to the general public.

    Lastly, radio observation is also preferred for low-budget users. 
    For optical observation, the price of optical telescope is typically at order of magnitude of $\$1000$ (all the price is in unit of US dollar throughout the paper) in order to produce fairly good image quality. Not to mention the extra expense on the automatic-tracking system for long exposure imaging, and the expense on optical filters, which significantly rise the entrance bar for hobbyists.
    In contrary, radio observation can be fairly cheap.
    As we will detailed in the later section of the paper, we manage to build the radio telescope at a cost of $\$150$ only, an order of magnitude cheaper than is optical telescope. 
    The significant price reduction lowers the entrance bar for hobbyists, and thus suitable for promoting astronomy to general public in an accessible way.
    
    \subsection{Experiment Design}
    Thanks to the recent development of software defined radio (SDR), radio observation is now much more accessible by general citizens. 
    SDR is an integrated instrument, that can convert the radio signal to data stream that can be readout using standard USB output to a computer.
    The price of SDR can range from \$40 to a few hundreds (for hobbyists), and we will demonstrate that in this paper, even a \$40 SDR is capable for measuring the diffuse emission from the Milky Way.

\begin{figure}
	    \centering
	    \includegraphics[width=\textwidth,angle=270,origin=c]{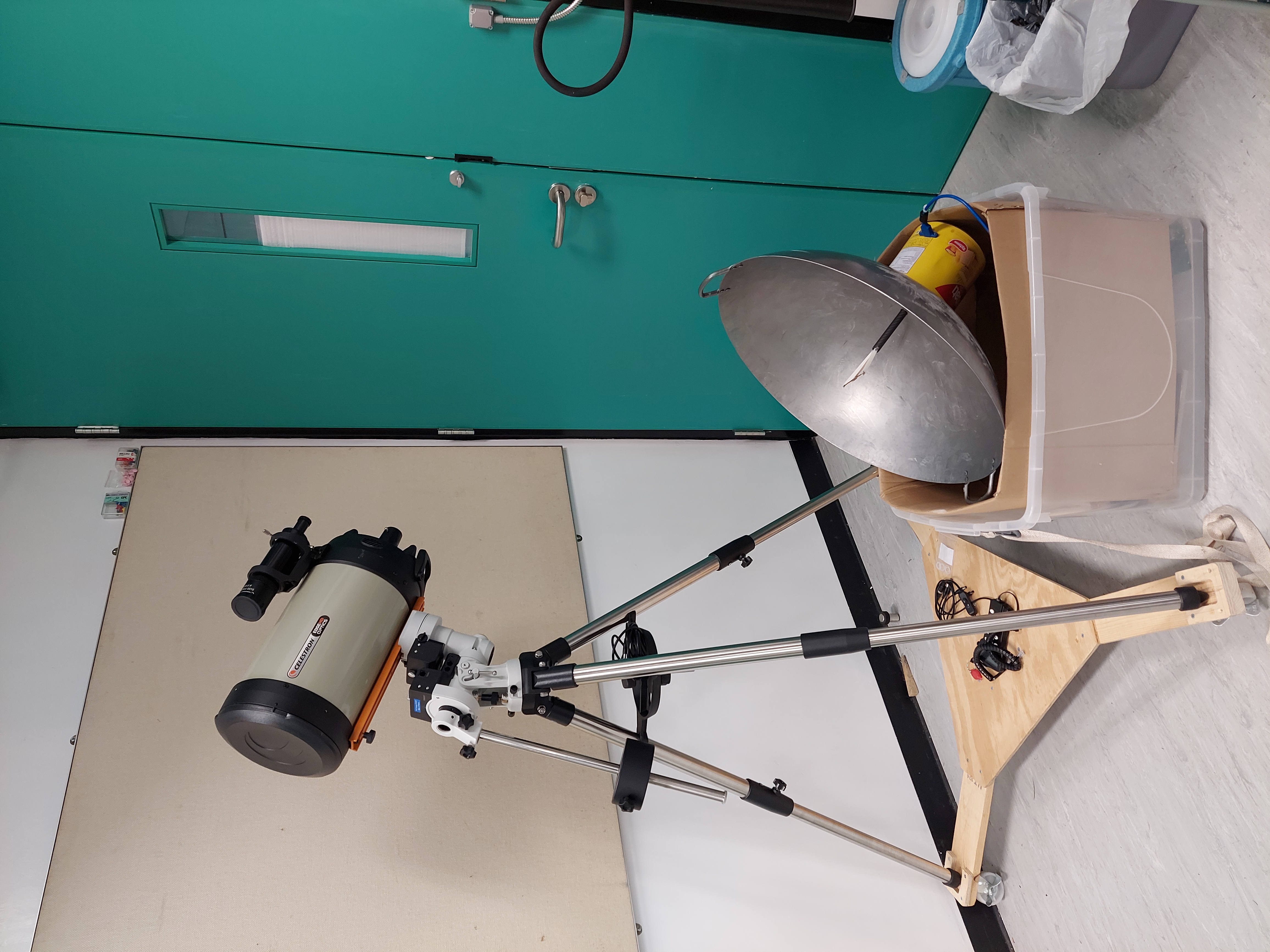}
	    \caption{Overview of the $61.0$ cm aperture WTH radio telescope (bottom right), compared with a $20.3$ cm Schmidt-Cassegrain optical telescope along with the mount from \textit{Celestron} (top left). The WTH system is temporally `mounted' on a plastic storage box with wheels for easier transporting the system to observation site.} 
	    \label{fig: wok-overview}
	\end{figure}

    These lead to our proposal of a fairly cheap design of a radio telescope for educational purpose.
    This radio telescope are designed specifically for observation at $1.42$ GHz, the emission frequency of the $21$ cm forbidden transition from neutral hydrogen.
    The telescope's gain profile is optimized at $1.42$ GHz, but is also capable to image neighboring wavelengths despite a drop in sensitivity.
    We specifically choose the readily accessible ingredients for constructing the radio telescope.
    Beside the electronic components that can be easily purchased online, all other hardware can be found in a typical kitchen.
    The main reflector of the telescope is a wok - a pan for stir-frying. 
    The wok is coupled with a dipole antenna made of copper wire, with its dimension optimized for $21$ cm observation.
    The readout electronics are shielded by a metallic cookie box, so as to prevent any noise picked up by the readout equipments.
    We name the system as Wok the Hydrogen (WTH).
    We show the photo of WTH system compared with a regular $20.3$ cm optical telescope in Fig.~\ref{fig: wok-overview}.

    We note that there are many previous implementations of amateur radio telescopes targeting $21$ cm emission.
    Those designs usually adopted a commercial satellite dish as the antenna. \cite{amateur-radio-dish-japan, amateur-dish-radio}
    The commercial satellite dish is usually customized either for WIFI transmission ($2.4$ GHz) or satellite transmission ($\sim 500$ MHz), and is therefore not optimal for $21$ cm hydrogen line detection ($1.42$ GHz).
    Furthermore, commercial satellite dish are often designed with a relatively higher focal ratio $f/D$, as the sources for the standard application of those dishes are fairly strong. 
    In contrary for radio astronomy, the source is weak and lower $f/D$ would be beneficial for the reduced pickups from side lobes.
    A more documented and well established project lead by Haystack Observatory, Massachusetts Institute of Technology \cite{mit-haystack-srt}, also outlined a more sophisticated dish design for education use, and the system can be extended to an intensity interferometer.
    There are also community projects that design horn antenna optimized for $21$ cm detection. 
    However, the physical size of a horn antenna is larger than a wok antenna for achieving similar gain characteristics.
    These projects are well documented in \cite{amateur-radio-horn-harvard,community-project-doc,physicsOpenLab-project, amateur-radio-horn-ieee, amateur-radio-horn-india}. (And the above list is far from a full list.)
    We also found that most of these projects are performed in a fairly radio quiet area instead of noisy urban environment, so that extra noise shielding are not necessary in those designs. 
    In our WTH project, we aim to address these issues. 
    Particularly, we detail the design of the antenna and documented the underlying engineering considerations, with a focus on how to mitigate the noise. 
    We also detail the data analysis pipeline, and the dark frame calibration procedure to clean up the data, which are necessary to improve the SNR in noisy environment.

    \subsection{Scientific Background}
    Historically, Rubin \cite{rubin-galaxy-rotation-curve} measures the recessional velocity of the stars in the Milky Way (MW), and find excessive velocity than is allowed by mutual gravitation among the stars alone. 
    This is usually quoted as the evidence of dark matter.
    
    The structure of MW still attracts huge scientific interest in the modern days. 
    The MW, being the galaxy that host the Earth, is the archaeological site for studying galaxy formation.
    The ability of resolving individual stars inside MW enable accurate dating of the star formation history in galactic environment.
    In relation to the motion of neutral hydrogen in the galactic plane - the observation target of our WTH project, different galaxy formation mechanism, and the structure of the dark matter halo can modify the phase space distribution of test particles (stars or gas particles), and therefore modify the corresponding velocity distribution.
    The kinematics of the stars inside the galaxy would therefore provide hints on studying the nature of dark matter. \cite{waveDM-compactibility-mw-satellite,wavedm-heating-dwarf,dynmaical-friction-in-waveDM-analytic,stellar-kinematics-waveDM,stellar-kinematics-dm-halo,stellar-streams-gaps-statistics-analytic}

    In principle, upon significant improvement of the system's noise figure, the system can be used to probe extragalactic $21$ cm emission.
    Such observation would also require a narrow beam to better avoid the galactic foreground.
    Considering the universe is made up of $\gtrsim 70 \%$ hydrogen, accurate survey of the distribution of hydrogen in the universe would provide valuable information about cosmology. 
    Extragalactic $21$ cm tomography 
    \footnote{Note that most $21$ cm tomography experiments are performed using radio interferometer, instead of single dish radio telescope as we implemented here. The use of interferometer not only increases the effective aperture of the telescope, but also allow higher resolution for constraining the small scale fluctuations of $21$ cm intensity.} 
    is currently one of the standard approach to study the matter distribution in the universe on the large scales $\mathbf{r} \sim 100$ Mpc. 
    There are several research-class radio experiments targeting the $21$ cm emission from the early universe \cite{chime-overview,chime-21cm-result}.
    There is also ongoing implementation of an upscaled design for $21$ cm detection \cite{chord-proposal}.
    The spatial distribution of the $21$ cm emission can provide insight on the distribution of cold gas, as its spatial distribution is tightly correlated with the distribution of dark matter. 
    Probing the sudden absence of $21$ cm emission from the primordial universe is also an active research direction, as such absence of emission are caused by cosmic reionization, where most neutral hydrogen are ionized by the first generation of stars. \cite{edge}

    \subsection{Educational Application}
    The WTH system can be utilized in various educational scenarios, catering to students ranging from elementary school kids to undergraduates. 
    For students with minimal mathematical background, the focus can be on using the WTH system to observe and collect real-time data. 
    The subsequent data analysis can be treated as a black box, allowing students to primarily emphasize the astronomical implications in their learning activities.

    On the other hand, students with a suitable mathematical and engineering background can delve into implementing both the hardware and software components of the telescope. 
    This project not only encompasses astronomy but also serves as an excellent application for digital signal processing and radio-frequency electronics. A solid understanding of Fourier transform is crucial for correctly reducing the data collected by the radio telescope. 
    Additionally, engineering practices in radio-frequency, such as impedance-matching, shielding, and signal balancing, significantly impact the system's sensitivity and are essential for the successful detection of the hydrogen signal.
    
    \subsection{Organization of the Paper}
    The paper would be organized as follows.
    We present and detailed the design and implementation of the hardware setup in Section \ref{sect: hardware}. 
    We then outline the mathematical formalism required to read the digitalized data, and the relevant data compression strategy in Section \ref{sect: software-overview}. 
    The observational test of the WTH project is documented in Section \ref{sect: observation}. 
    The data pipeline for extracting astronomical parameters from the noisy, compressed data are presented in Section \ref{sect: data-pipeline}.
    We finally conclude in Section \ref{sect: conclusion}.
    
    \section{Hardware Setup} \label{sect: hardware}
\begin{figure}
	    \centering
	    \includegraphics[width=\textwidth]{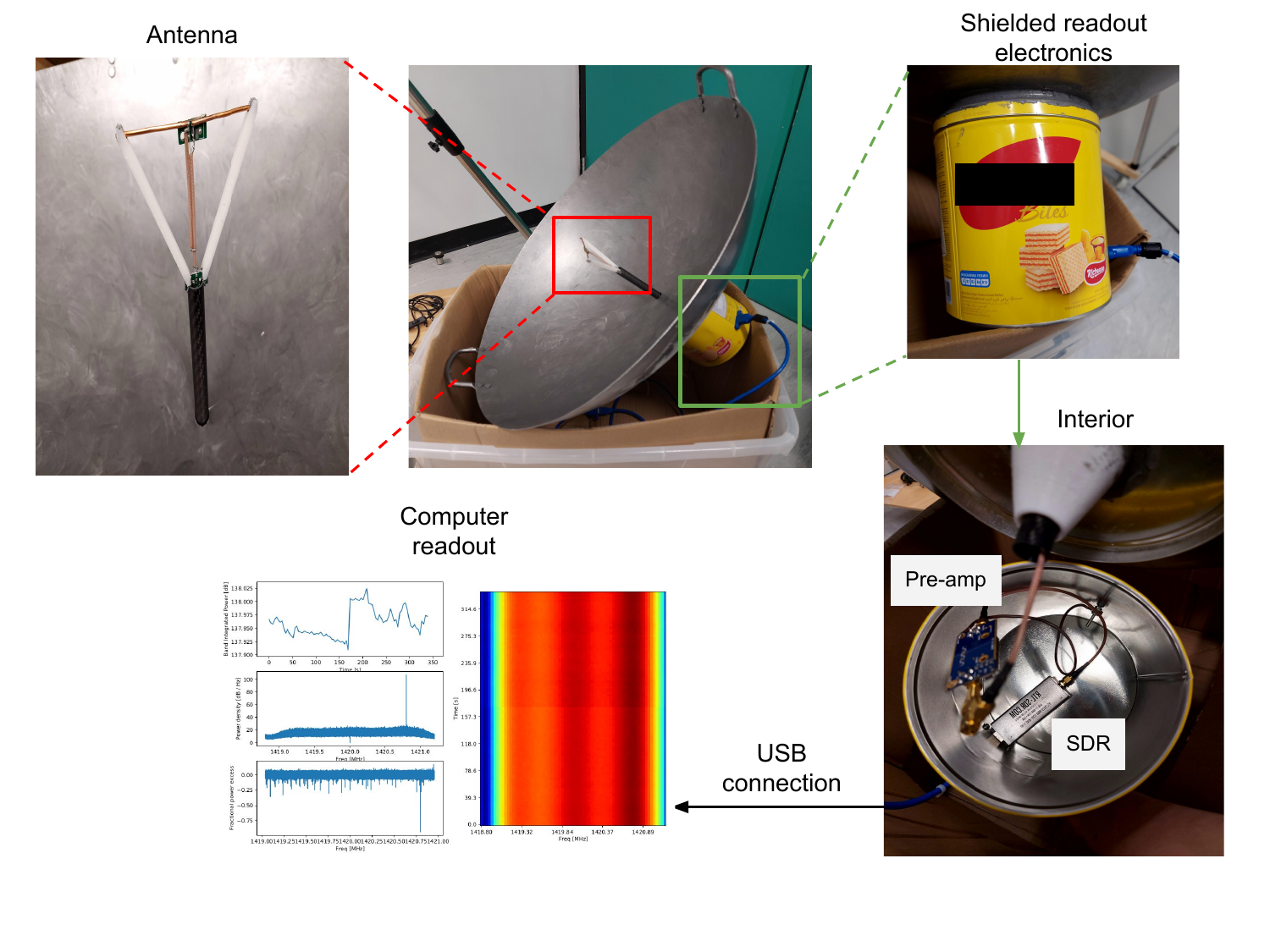}
	    \caption{Exploded view of the radio telescope hardware implementation. The user interface of the readout softwareis detailed in Section~\ref{sect: software-gui} and Fig.~\ref{fig: readout-gui-demo}. The design of the antenna is explained in Fig.~\ref{fig: dipole-antenna}.}
	    \label{fig: system-diagram}
	\end{figure}

    The overview of the system is shown in Fig.~\ref{fig: system-diagram}. 
    In the following subsections, each of the components would be explained in detail.
    
    \subsection{Dish - Dipole Antenna Design}
    \subsubsection{The Wok Reflector}
    For educational initiatives, we aim to construct a 21-cm line antenna that is simple, lightweight, and cost-effective. The ideal criteria for the antenna include a price tag under a hundred US dollars, a weight less than $10$ kg to facilitate single-person deployment, and a design that doesn't demand extensive machining. 
    It is also essential that the antenna is durable enough for multiple experiments and self-standing so that no external support is necessary.

    Historically, hobbyists constructing small-scale 21-cm radio antennas have predominantly employed the horn and waveguide design \cite{amateur-radio-horn-harvard,community-project-doc,physicsOpenLab-project, amateur-radio-horn-ieee, amateur-radio-horn-india}.  
    This design can be realized using thick aluminum boards, substantial enough to be self-supporting, or even cardboard combined with aluminum foil. However, there are challenges: acquiring large aluminum boards can be costly, particularly in Hong Kong. 
    Thick aluminum also demands more intricate machining processes. On the other hand, the cardboard-aluminum combination is not durable in the humid climate of Hong Kong. Hence, we are exploring alternative designs.

    Modern radio observatories predominantly use the dish reflector design, specifically parabolic, over the horn as their main reflector. This shift is attributed to the cost-effectiveness of dishes compared to larger horns and their broader acceptance bandwidth. 
    While our setup isn't large-scale, we draw inspiration from the modern parabolic dish reflectors, particularly the deep dish reflector (f/D $\sim 0.21$) design seen in the CHORD project. \cite{chord-proposal}  
    We see a deep dish reflector as reminiscent of a large bowl. This similarity leds us to explore kitchenware for potential modifications into a radio reflector.
    
    Most kitchen utensils weren't suitable as they generally have flat bottoms for stability, which isn't ideal for radio reflection. However, we identified that the geometry of a commercial Chinese wok closely aligns with a small dish reflector. These woks, primarily intended for commercial use, are large, deep, crafted from iron sheets, feature a round bottom, and are lightweight. All these features align with our requirements. 
    Consequently, we purchased a 61cm Chinese wok from a local outlet for approximately $30$ USD to conduct tests. Prior to finalizing our antenna design, we need to ascertain its geometry and focal point, which will be discussed in the subsequent section.
    
    \subsubsection{Focal Point Estimation}
    Consider a wavefront from infinity scatter off a reflector. 
    Depending on the geometry of the reflecting surface, the reflected wavefront across the surface can be focused onto the same location, and the resulting wavefront can either be in phase or out phase.
    In this subsection, we will show how to ensure constructive interference occurs - which defines the focal point, so that the measured signal strength can be maximized.

    The determination of the focal point can be drastically simplified if the reflecting surface is parametric. In particular, a radio reflector dish is typically described by conic section functions: spherical, parabolic, or hyperbolic surface. 
    Here we will consider the two most common and simple cases: spherical and parabolic shapes, which should approximately describe our wok's shape. We measured the diameter of the wok to be $61.0$ cm, and the depth to be $17.8$ cm.

    For a spherical concave mirror, the focal length of a spherical mirror is one-half the radius of curvature of the mirror. Here the radius of curvature of the wok can be simply calculated to be $35.0$ cm, derived from the diameter and depth of the wok. Thus, hte focal point $h_f$ lies $17.5$ cm from the bottom of the wok, or roughly on the surface of the opening.  

    For a parabolic surface, the calculation of focal length is even simpler: $h_{f} =\frac{r^2}{4d}$, where $r$ refers to the radius and $d$ refers to the depth of the wok.
    Hence the calculated focus distance on our wok is $h_f \sim 13$ cm. 
    Notice that the parabolic and spherical focal spot differs quite a lot.  
    We outline the attempted determination of the geometry of the wok in Appendix~\ref{app: wok-geometry}, and found that we cannot draw conclusion on the geometry of the wok. 

    Lastly, we have to determine the center $x_f$ of the wok. 
    To do this, we first align the wok to be parallel to the gravitational acceleration, 
    and such alignment is ensured by a level ruler.
    Afterwards, we utilize the curvature of the wok, to slide a marble down to the wok center.
    The equilibrium position of the marble would therefore indicate the center $x_f$ of the wok. 
    This center $x_f$ determination process is repeated, with different initial conditions (both position and velocity) of the marble for a few times to ensure convergence.

    Although the above calculation provide a determination of the geometrical focal spot of the system, the actual focal point for optimal gain is not exactly at the geometrical focal spot. 
    As we will show in Section.~\ref{sect: gain-simulation}, the optimal gain focal spot is $h_f^*  = 15.0$ cm, which compensates for the difference in the dielectric properties of iron (the wok) and the air. The wok is constructed according to the simulation, and the adopted design parameter for the wok is shown in Table.~\ref{tab: antenna-specification}.
   \subsubsection{Dipole Antenna}
    An antenna element should be placed in the focal spot of the wok reflector to convert EM waves into electric current. A dipole antenna is one of the simplest and most widely used types of antennas which gives an omnidirectional radiation pattern to illuminate our deep wok. Since the dipole antenna is not sensitive to the axial direction, the illumination on the wok will be around $70\%$. 

    We build our dipole antenna on $2$ mm copper wires, which are strong enough to support itself from bending. 
    \footnote{Copper wire can also be found in kitchenware, for example by desembling from microwave oven.} 
    Ideally, each arm of the dipole antenna should have a length of $\lambda/4 = 52.75$ mm $\approx 53$ mm, with an infinitesimal gap between two arms, so the overall length equals $\lambda/2$. Practically, the gap needs to be large enough for soldering and avoid shorting, so the arm length will be slightly shortened; furthermore, we will fine-tune the arm length to provide the best sensitivity and impedance matching with the wok reflector. 
    The finally adopted design parameter is shown in Table.~\ref{tab: antenna-specification}.
    
    Naturally, a free-standing dipole antenna in air produces an impedance of $(73+43j)\ \, \Omega$. By coupling the dipole antenna to our wok reflector, we will match the overall antenna impedance to the readout system input impedance of $\approx (50+0j)\ \, \Omega$.
    
    \subsubsection{Gain and Impedance Simulation} \label{sect: gain-simulation}
\begin{figure}
	    \centering
	    \includegraphics[width=0.7\textwidth]{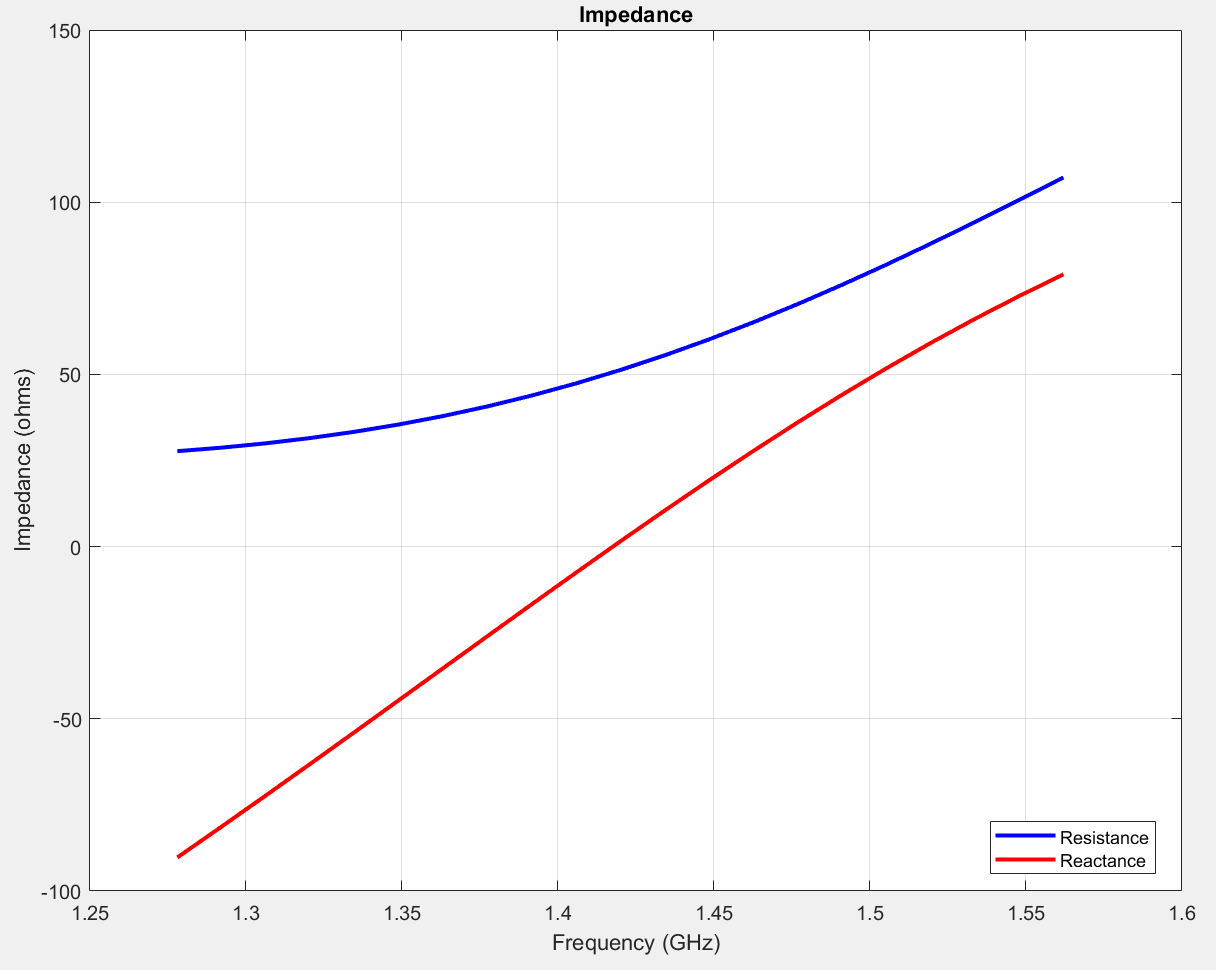}
	    \caption{Simulated impedance of the reflector-antenna system, from $1.3$ GHz to $1.55$ GHz. } 
	    \label{fig:impedance}
    \end{figure}

\begin{figure}
    \centering
    \begin{subfigure}[b]{0.45\textwidth}
        \centering
	    \includegraphics[width=\textwidth]{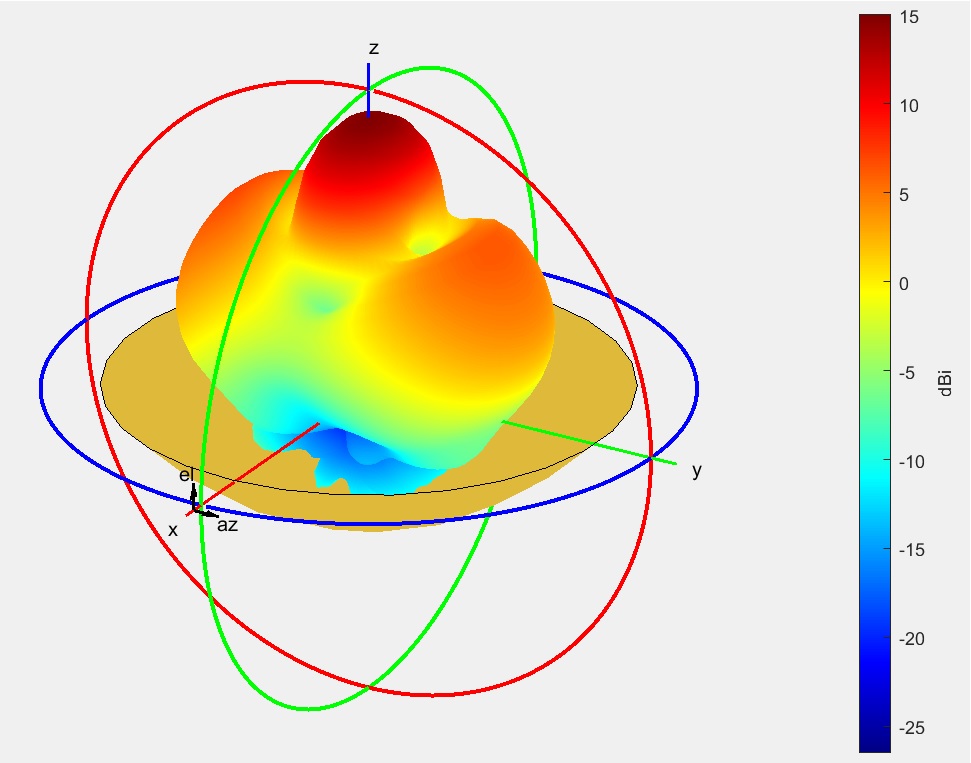}
	    \label{fig: beam_pattern_3d}
    \end{subfigure}
    \hfill
    \begin{subfigure}[b]{0.45\textwidth}
	    \centering
	    \includegraphics[width=\textwidth]{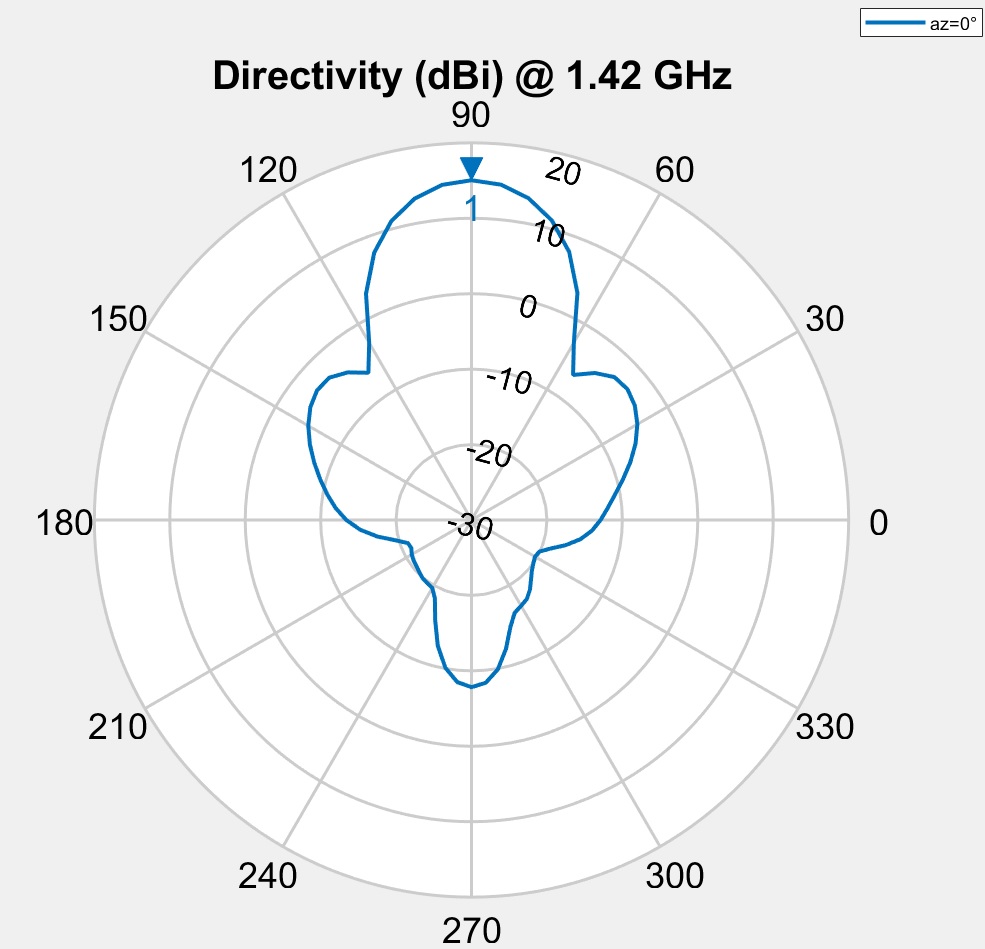}
	    \label{fig: directivity}
    \end{subfigure}
        \caption{Simulated gain characteristic of the radio telescope. Left panel: the 3D gain profile; Right panel: the projected directivity profile along the section that cut the plane of highest directivity gain (the red ring on the left). }
        \label{fig: beam}
    \end{figure}

    Following the numbers calculated above, we optimized the antenna and reflector geometry through the simulation with \textsc{matlab} Antenna Toolbox. We simulated the geometry under the assumption of a spherical wok, while changing to a parabolic wok will give minimal disparity (directivity $\pm 0.5$ dBi, impedance $\pm 2 \, \Omega$). 
    The simulated impedance profile is shown in Fig.~\ref{fig:impedance}. We optimized the dipole antenna length to $9.7$ cm instead of $\lambda/2 = 10.55$ cm for matching impedance to $(52+0j)\, \Omega$ under the focal length of $15$ cm. 
    Further reducing the focal length will push impedance to $(50+0j)\, \Omega$, therefore minimizing the return loss, but at the cost of reducing directivity due to loss in the illumining area. 
    The simulated beam pattern and directivity are shown in Fig.~\ref{fig: beam}, suggesting that our antenna design can provide a directivity of $15.7$ dBi in ideal conditions, and a FWHM beam width of $24^\circ$. The finalized antenna specifications are listed in Table.~\ref{tab: antenna-specification}.

    \begin{table}
        \centering
        \begin{tabular}{c|c}
            Reflector Diameter & $61.0$ cm ($24.0$ inch) \\
            \hline
            Reflector Depth & $17.8$ cm ($7.0$ inch)\\
            \hline 
            Dipole antenna total length & $9.7$ cm \\
            \hline
            \hline
            Adjusted focal length$^{*}$ & $15.0$ cm \\ 
            \hline
            Beam width\textsuperscript{\textdagger} &  $24^\circ$ \\
            \hline
            Directivity & $15.7$ dBi \\
            \hline
            Impedance & $52+0j$ $\, \Omega$ \\
        \end{tabular}
        \caption{Specification of the reflector-antenna system. 
        The first half of the table are measured quantities, while the second half (below the double horizontal lines) is derived from simulation. 
        Some remarks on the table follows. 
        $^{*}$: This adjusted focal length is defined as the optimal focal length after adjustment from impedance matching, and is slightly different from the geometrical focal length. \textsuperscript{\textdagger}: The beam width is defined as the width of the $-3$ dBi region (around the maximum gain direction) in the main lobe. The numerical choice of $3$ dBi is to align with the usual definition of full-width half maximum. 
        }
        \label{tab: antenna-specification}
    \end{table}

    \subsection{Signal Processing Chain}
    \subsubsection{Signal Balancing}
     A dipole antenna is balanced, as the two arms of a dipole antenna are symmetric, and each carries half the strength of the received signal. However, most readout systems take only asymmetric input, i.e. input wire with reference to ground. If we connect the symmetric antenna to an asymmetric load (readout system input), half of the signal will be lost to the ground, and reflection from the ground will also affect our measurement accuracy. 

\begin{figure}
        \centering
        \begin{subfigure}[b]{0.45\textwidth}
            \centering
    	    \includegraphics[width=\textwidth]{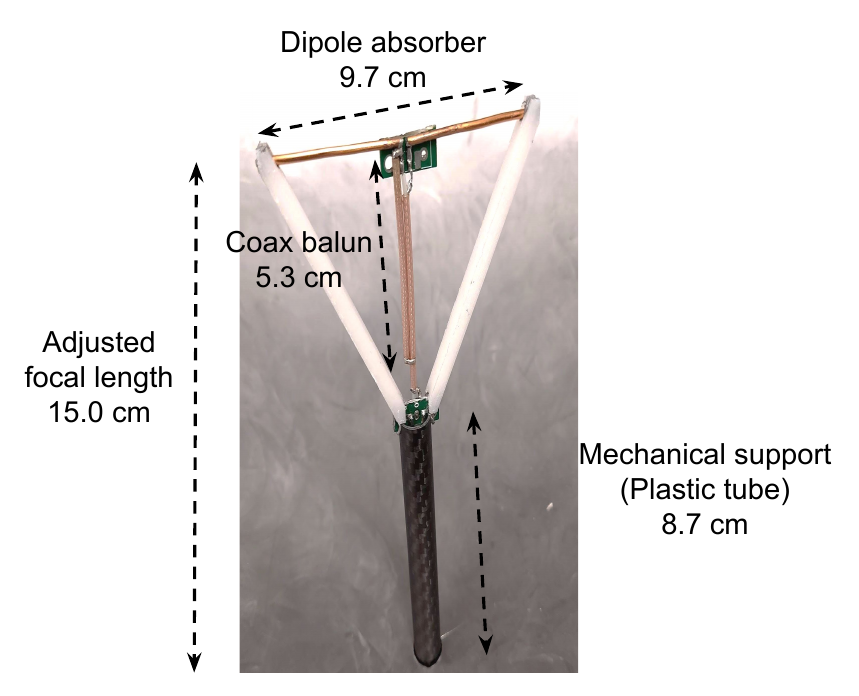}
        \end{subfigure}
        \hfill
        \begin{subfigure}[b]{0.45\textwidth}
    	    \centering
    	    \includegraphics[width=\textwidth]{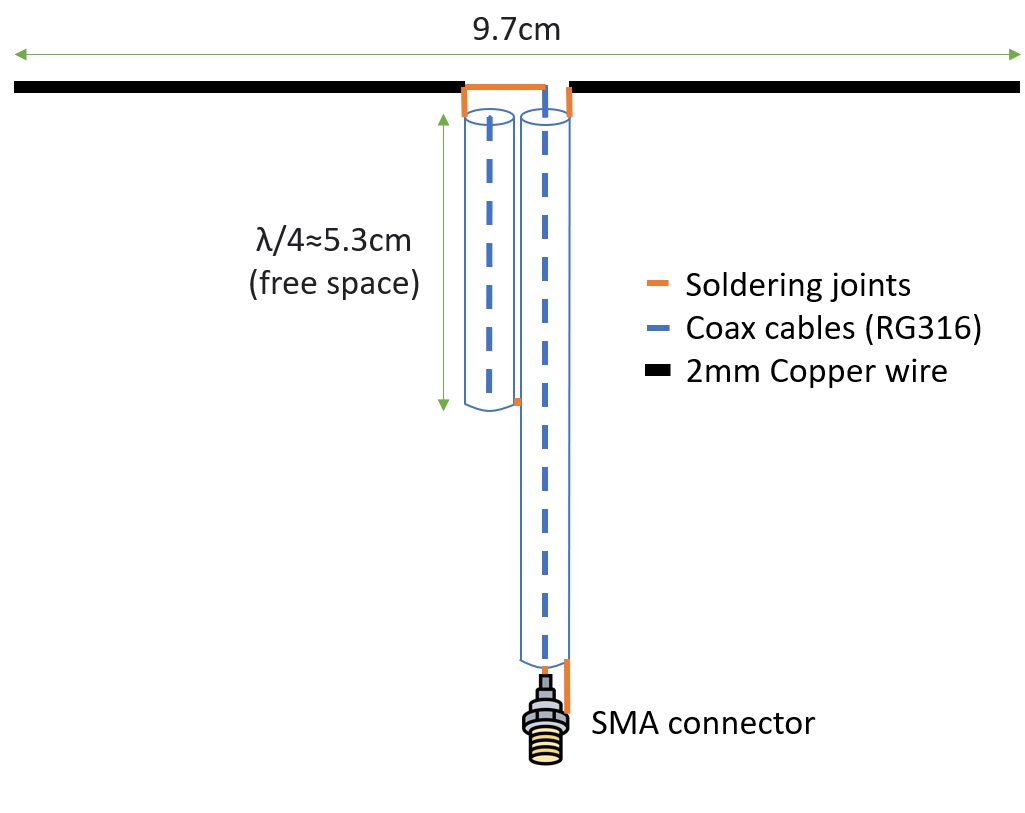}
        \end{subfigure}
        \caption{The configuration of the dipole antenna. Left: the mechanical configuration of the dipole antenna. The white diagonal structure is just to provide mechanical support to the dipole antenna, so as to prevent the deformation of the copper wire dipole. Right: The schematic of the antenna feed system, with emphasis on the folded quarter-wave balun.}
	    \label{fig: dipole-antenna}
	\end{figure}

     Therefore, we employed a simple folded quarter-wave balun (balanced - unbalanced converter) design, as shown on the right panel in Fig.~\ref{fig: dipole-antenna}. Such design provides a 1:1 impedance conversion (i.e. from $50\, \Omega$ balanced to $50\, \Omega$ unbalanced) through two parallel coax cables with low loss and compact design. 
     
    \subsubsection{Pre-amplifier}
    After the dipole antenna absorbed the incoming radio wave, the signal would flow along the signal line for a short distance (i.e.: the adjusted focal length of the wok).
    We put our first stage amplification immediately behind the wok so as to minimize the cable loss.
    The first stage amplification system is the commercial product \textit{Nooelec SAWBird+ H1} specifically designed for $21$ cm detection. 
    This composite low noise amplifier (LNA) first includes a surface acoustic wave filter to isolate the signal at frequency $f = 1408_{-32}^{+33}$ MHz.
    This prevents the unwanted pickup of noise that mixes with the target signal in the readout system.
    The filtered signal would then enter the amplification module which provide a gain of {$42$ dB} to the $21$ cm emission. 
    This number is quoted from the datasheet \cite{sawbird-amp-datasheet}, and we have verified that the gain is in good agreement with the datasheet. 

    The operation voltage of this amplifier is $3$-$5$V, which matches the voltage standard of USB-driven power supply.
    The LNA is powered by a dedicated bias-T design, so that the DC power share the same interface with the RF output. 
    With this configuration, the LNA can draw DC power indirectly from the USB interface of the computer via the software defined radio (as detailed in the next subsection).
    This is optimal to prevent the mild instability in the DC power to contaminate the signal via secondary path.
    
    \subsubsection{Readout: Software Defined Radio (SDR)}
    After the pre-amplification stage, the signal is directed to an integrated commercial readout system, the \textit{RTL-SDR 3.0} software defined radio (SDR). \cite{rtl-sdr-datasheet}
    This SDR is a composite system, based on the \textit{R820T2} chip manufactured by Rafael Microelectronics. 
    It consists mainly of a tunable local oscillator, that allow mixing and down conversion of the signal to the baseband of bandwidth $2.4$ MHz. 
    Part of the signal is also phase shifted by $90^\circ$ to implement the in-phase and quadrature (I/Q) sampling. 
    After such pre-processing, the built in analog-to-digital convertor (ADC) would convert the time variation of baseband voltage into $8$ bits bytestring, and output the digitalized data stream via USB interface. 
    The digitalized data stream is then post-processed by a computer, as would be detailed in Section.~\ref{sect: data-pipeline}.

    As a remark, this SDR also includes a bias-T module, which is utilized to power the LNA as mentioned in the last subsection.
    
    \subsubsection{Noise Mitigation}
    The aforementioned signal path, although simple, can still be susceptible to substantial noise entering through the secondary path. 
    In this subsection, I will briefly describe the strategies we have implemented to alleviate this noise.

    One of the initial measures we have taken to prevent the signal wire from acting as an antenna and absorbing unwanted noise is to use short coaxial cables throughout the primary signal path. 
    Additionally, all amplification stages are shielded within a metallic box, as detailed in Section~\ref{sect: shielding}.
    
    Despite these precautions, residual noise can still be present. To address this, we use additional wire to ground all the electronic components into a common ground. 
    This helps to mitigate the problem of ground loops, where slight potential differences across the ground terminals can establish secondary paths for distorted signals to propagate within the system.
    
    Once the signal is digitized in the software-defined radio (SDR), it is sent to a computer for further signal processing. 
    To connect the SDR to the computer, we use an unshielded USB extension cable that is placed outside the shielding box. However, we have observed that this lengthy USB cable can introduce structured and time-varying noise into the system. 
    This noise is likely caused by the radiation of the digital signal on the USB cable, which is then picked up by the antenna frontend. 
    To mitigate this issue, we have installed several ferrite chokes along the USB extension cable. We have found that this significantly reduces the temporal variation in noise.
    
    \subsection{Mechanical Design}
    \subsubsection{Antenna Holder}
    We place two diagonal plastic rods to prevent the deformation of the copper wire dipole antenna. These plastic rods are made on small, household 3D printers based on white PLA plastics. The whole antenna is placed on the focal spot by a plastic pipe with tunable length (it can be pulled/pushed to adjust antenna feed height). The connection SMA cable is hidden inside the tube, directly tunneling into the shield box.

    \subsubsection{Shielding Box} \label{sect: shielding}
    A metallic cookie box as shown in Fig.~\ref{fig: system-diagram} is used to enclosed all the electronic devices in the WTH system.
    The cover of the cookie box is stuck on the back of the wok by Permatex 59903 Ultra Grey Gasket Maker with 3D printed PLA plastic fillings.
    We drilled an opening on the cover to allow the coax connected to the antenna to immerse into the box, and an opening on the side to allow the USB cabling to the computer readout. 
    We connect the box to the ground terminals of all the electronic devices inside the box, so as to avoid ground loop noise.
    We have verified that this setup, albeit being simple, is very effective in preventing the wiring between electronic components to pick up any noise outside the system.
    The noise floor is suppressed by more than $30$ dB after using this proper shielding design.
    
    \subsubsection{Temporary Mounting}
    Measuring the faint signal from galactic neutral hydrogen requires a long exposure time. Additionally, if the reflector system comes into contact with other charged materials, extra noise can be induced. 
    \footnote{We have verified that manually holding the wok increases the band-integrated power received by the system by approximately 10 dB, which in turn raises the noise floor by +10 dB.}
    These concerns emphasize the importance of using a proper holder to mount the system.

    Instead of designing a custom-made mounting system, we have found that a typical plastic storage box (as shown in Fig.~\ref{fig: wok-overview}) with an opening slightly smaller than 61 cm is suitable for mounting the wok. 
    The hollow interior of the storage box is compatible with the spherical shape of the wok, allowing for adjustment of the wok's angle of elevation. Although part of the wok may be below the plastic box when aligned horizontally, the blocking effect is minimal considering that plastic is nearly transparent to radio wavelengths. There may be some mild modification in the gain characteristics due to the slightly different dielectric properties of plastics compared to air. 
    Finally, the wheels underneath the plastic storage box enable easy transportation of the radio telescope to the observation site.
    
    Since the mechanical support of such a mount is concentrated on a few contact points between the wok and the plastic box, the wok may deform. To minimize this deformation, we align the two holders of the wok vertically (as shown in Fig.~\ref{fig: wok-overview}). 
    Additionally, a cardboard box is placed inside the plastic storage box. As cardboard is more susceptible to deformation, it helps alleviate the mechanical stress applied to the wok, thus reducing the possibility of reflector deformation.
    
    Lastly, counterweights are placed inside the plastic box to address the issue of the heavy metallic wok shifting the center of mass of the entire WTH system, which would make the system unstable for balancing the torque. 
    In practice, we place other testing equipment and accessories inside the plastic box for convenience. These loads also serve as the required counterweights to stabilize the system.

    \section{Software Setup} \label{sect: software-overview}
    \subsection{Data Stream Format}
    The SDR measures the voltage across the antenna at a sampling frequency of $f_s = 2.4$ MHz. Furthermore, at each single time step $t$, the input voltage is splitted into 2 channels, with one of the channels measure the voltage after $\Delta \phi = 90^\circ$ phase shift.
    The two channel measurement of voltage in each time step therefore form a I/Q sample of voltage $V(t) = V_0(t) + j V_{90}(t)$. 
    The voltage measured $V_0, V_{90}$ at each sample time step are then digitalized into $8$ bits integer each, therefore feeding data at a rate of $2 \times 2.4$ MHz $= 4.8 $ MB/s, equivalently generating $\sim 17$ GB of data every hour. 

    With this sampling rate, the spectral resolution would be of $\Delta f = 1$ Hz, which is very small compared to the line width of the $21$ cm neutral hydrogen that can be of $\Delta f_{\rm 21} \approx 10^5$ Hz. 
    It should be hence obvious that some form of data compression can be done to minimized the disk space required for data storage.
    \subsection{Data Compression}
    In the application of measuring the spectral emission from $21$ cm neutral hydrogen, there is redundant information in the I/Q sample time series. 
    In particular, we are interested in the frequency spectrum of the associated I/Q sample. 
    The frequency domain of I/Q samples is defined via the discrete Fourier transform (DFT): \footnote{Throughout the paper, we use the conventional notation $j=\sqrt{-1}$ used in engineering.}
    \begin{equation} \label{eq: fourier-mode}
        V(f) = \sum_{i=0}^N V(t_{\rm ini} + \frac{i}{f_{s}}) \exp({j\frac{if}{N f_s}}),
    \end{equation}
    this complex number $V(f)$ determined by the weighted sum of the I/Q samples $V_0(t) + j V_{90}(t)$ can be expressed more convenient by the magnitude of the power spectrum $P(f)$:
    \begin{equation}
        P(f) \equiv | V(f) |^2 ,
    \end{equation}
    which has the physical meaning of the total power contained in the wave of frequency $[f, f+\Delta f_s]$.   
    The argument of the complex power spectrum is related to the relative phase across wave of different frequency, which is irrelevant for our application. 

    When we perform DFT and convert our raw I/Q samples into magnitude power spectrum, the complex samples are mapped to real numbers, effectively reduce the storage by a factor of $2$.  
    Apart from this conversion of I/Q samples into magnitude power spectrum in real time before storage, extra disk space can be saved by appropriate binning and averaging of different magnitude power spectra.

    We choose $N = 2^{15}$ for obtaining the DFT using Eq.~\ref{eq: fourier-mode}, which can be converted to the frequency resolution of $\Delta f = f_s/N = 36.6$ Hz. 
    After converting the DFT to magnitude power spectra in real time, we also stack the magnitude power spectra in a time window of $t_{\rm stack} \approx 4$ s, and store the stacked spectra subsequently. 
    This pre-processing procedure allows more optimized storage usage, and the compression rate can be adjusted by tunning $N$ and $t_{\rm stack}$. 
    The numerical value $t_{\rm stack} \approx 4$ s is a balance between the monitoring of temporal variation of noise and storage space, so that transient noise can be identified and masked during post-processing.
    
    \subsection{User Interface} \label{sect: software-gui}
    \begin{figure}
	    \centering
	    \includegraphics[width=\textwidth]{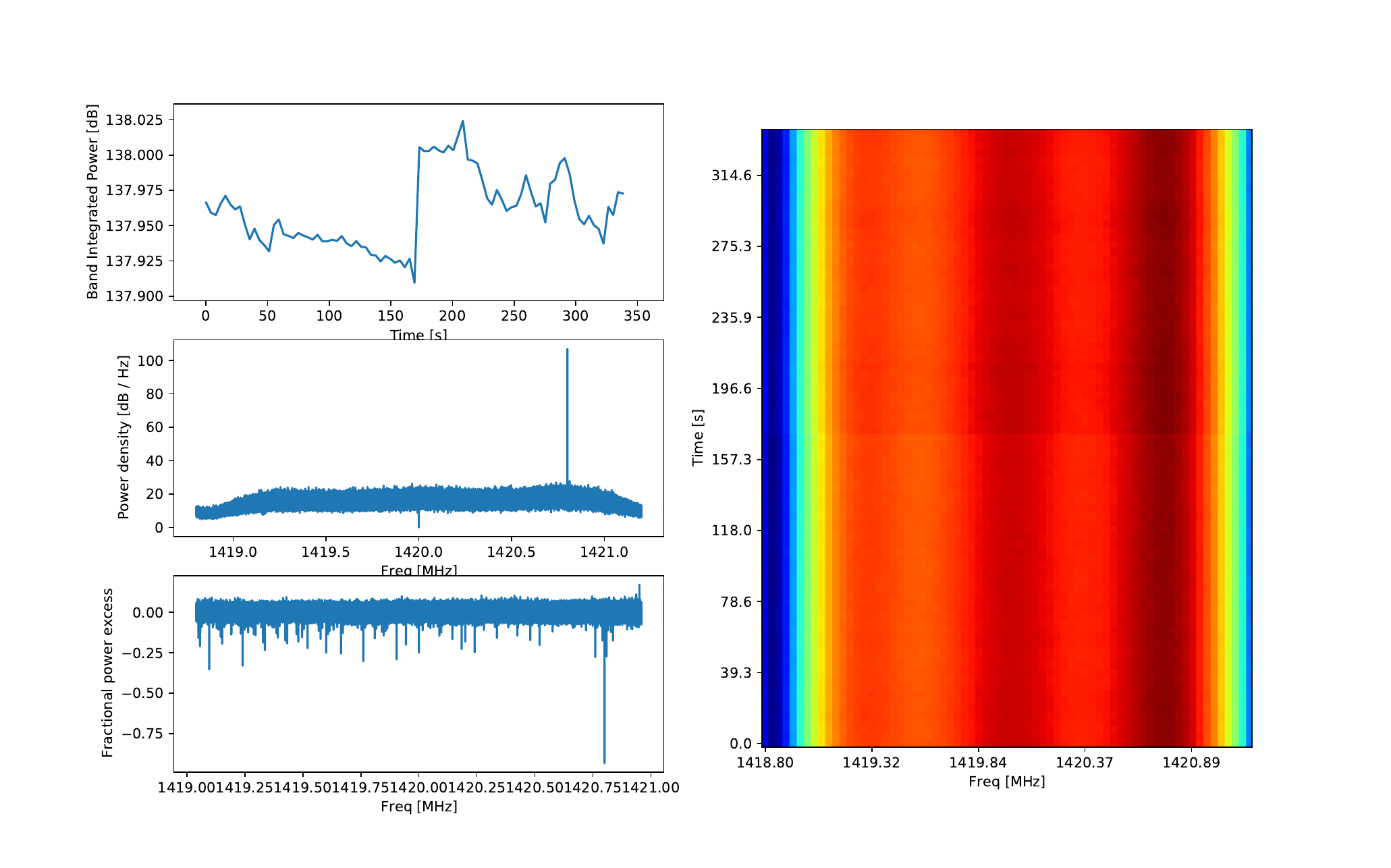}
	    \caption{An example real-time readout GUI for monitoring the state of the system. } 
	    \label{fig: readout-gui-demo}
	\end{figure}
    To facilitate trouble-shooting, the readout system would display the following diagnostics:
    \begin{itemize}
        \item The total (band-integrated) power received by the system. This is to identify if there is any transient noise during the course of obesrvation. On the other side, we find that the band-integrated power of the system are stabilized only after warming up the system for $\sim 10$ minutes. 
        \item The instantaneous power spectrum received by the system. This is an important diagnostic for the identification of any spectral feature in the environment.
        \item The time evolution of the spectrum (waterfall diagram). Line-alike noise transients, apart from contributing to a tiny change in the total power received, would leave a remarkable feature in the frequency domain.
        \item Real time stacked spectrum, accumulated since the begin of the observation. In additional to accumulation, there is a option for the real-time subtraction to a calibrated dark frame (detailed in Section~\ref{sect: dark-frame}). This facilitates the identification of any strong residual signal in the current pointing.
    \end{itemize}
    A snapshot of the user interface is shown in Fig.~\ref{fig: readout-gui-demo}. 
    The data shown on the figure is the actual data we obtained in the Sagittarius pointing as would be detailed in Sect.~\ref{sect: observation}.
    
    \section{Observation} \label{sect: observation}
    \subsection{Site Selection}
    To test the applicability of the system in noisy environment, we choose to run the observation in the campus of Hong Kong University of Science and Technology, located in Clear Water Bay, Hong Kong. (Whereby the authors are/was affiliated at during the experiment.)
    The university is surrounded by residential areas which contribute to significant noise due to human activities. 
    The campus is located $\approx 2$ km north-east of the Tseung Kwan O area, which is a residential area hosting a population of $0.4$ million. 
    At $\approx 5$ km north from the campus, there is the town Sai Kung which is a famous spot for tourism. 
    Along the $10$ km line of sight towards the west of the campus, there are many towns hosting a total population of $2$ million in the Kowloon area, with a population density of $\sim 43000$ km$^{-2}$.
    This number is to be compared to the lower population density in Toronto, the most populated city in Canada $\sim 28000$ km$^{-2}$. 
    Also noted that at $40$ km north-west of the site, there is the city center of Shenzhen, one of the most populated city in Southern China.

    Within the campus, we select the most radio quite possible area: near the water sports center, which is $500$ m away from the main academic building, but is only $100$ m away from the sports center.
    
    \subsection{Survey Strategy}
    Before starting the observation, we first run the system with no specific pointing to ensure the system stabilized. 
    We speculate that such short period of instability after starting of the system is due to the heat accumulated in the SDR. 
    The system is stabilized within $1$ dB of band-integrated power only after running the system for roughly $10$ minutes, possibly after the temperature of the SDR approach equilibrium, in which the heat generation rate by the SDR is in equilibrium with the heat loss rate. 
    
    After system stabilization, we start to take $10$ minutes dark frame by pointing the system to at least $30^\circ$ away from the galactic plane. 
    The data taking is real-time monitored by the GUI shown in Section~\ref{sect: software-gui} and Fig.~\ref{fig: readout-gui-demo}, so that transient radio signal can be immediately identified.

    Measurement of $3$ different galactic longitude is immediately followed after the dark frame taking. 
    Considering the $24^\circ$ beam width of the main lobe, precision pointing is not required. This allow us to point to the target by alignment with the nearby constellations along the galactic plane. 
    In particular, we choose to sweep through the galactic plane from the center of Milky Way, as can be identified by the constellation Sagittarius. 
    The pointings are specified in Table.~\ref{tab: pointing-specification}.
    The pointings are adjusted manually, with using a pair of binoculars to help the identification of the correct constellation in the environment with city light pollution.
    During the exposure of roughly $10$ minutes each, the system is left untouched, so that the system would scan different regions on the sky as the Earth rotates. 
    Again, we argue that this effect is unimportant considering the $24^\circ$ beam width.
    As the other region of the Milky way are still below the horizon at the time of observation, we only measure the radio emission in these few pointings.

    After the measurement of the galactic signal, we end the observation by taking an extra $10$ minutes dark frame, with a pointing different from the previous dark frame taken during the start of observation. 
    This ending dark frame is for verifying if any change of system's sensitivity and noise characteristics after the system have been operating for long.
    We described the analysis and the application of dark frame in Section~\ref{sect: dark-frame}. 
    
    \subsection{Description of Data}
    \begin{figure}
	    \centering
	    \includegraphics[width=.8\textwidth]{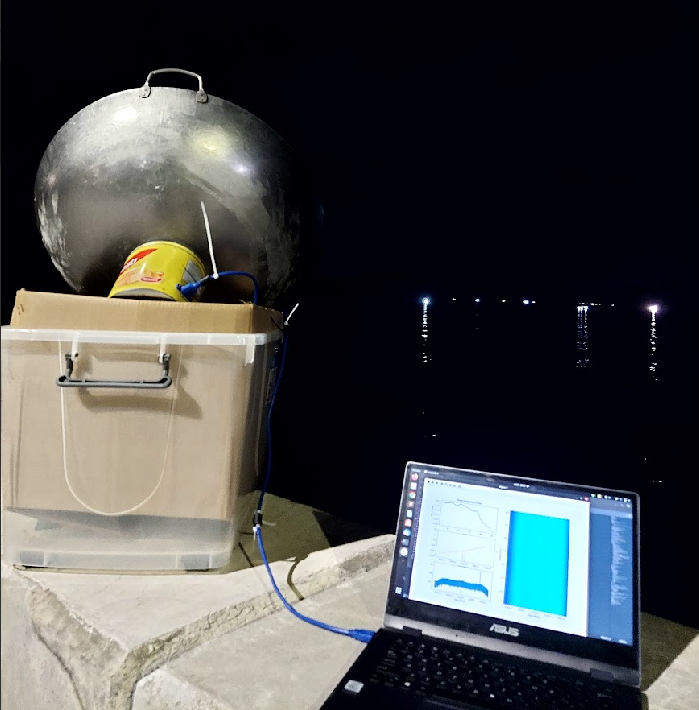}
	    \caption{The configuration of our setup during observation, along with a brief view of the observational environment. } 
	    \label{fig: wok-in-onbservation}
    \end{figure}

    \begin{table}
        \centering
        \begin{tabular}{c|c|c|c|c|c}
            Constellation & RA & Dec & galactic $\ell$ & galactic $b$ & Exposure time [s] \\
            \hline \hline
            Sagittarius & 270 & -30 & 0 & 0 & 346\\
            \hline
            Aquila  & 290 & 0 & 40& -10& 401\\
            \hline
             Cygnus & 300 & 40 & 80& 0& 322\\
             \hline
        \end{tabular}
        \caption{Summary of the pointings in the observation. Note the coordinate information are all in unit of degree, and are rounded to nearest $10^\circ$, as the beamwidth of the antenna is $24^\circ$. Note the coordinates are normalized to J$2000$ epoch. }
        \label{tab: pointing-specification}
    \end{table}

    \begin{figure}
	    \centering
	    \includegraphics[width=\textwidth]{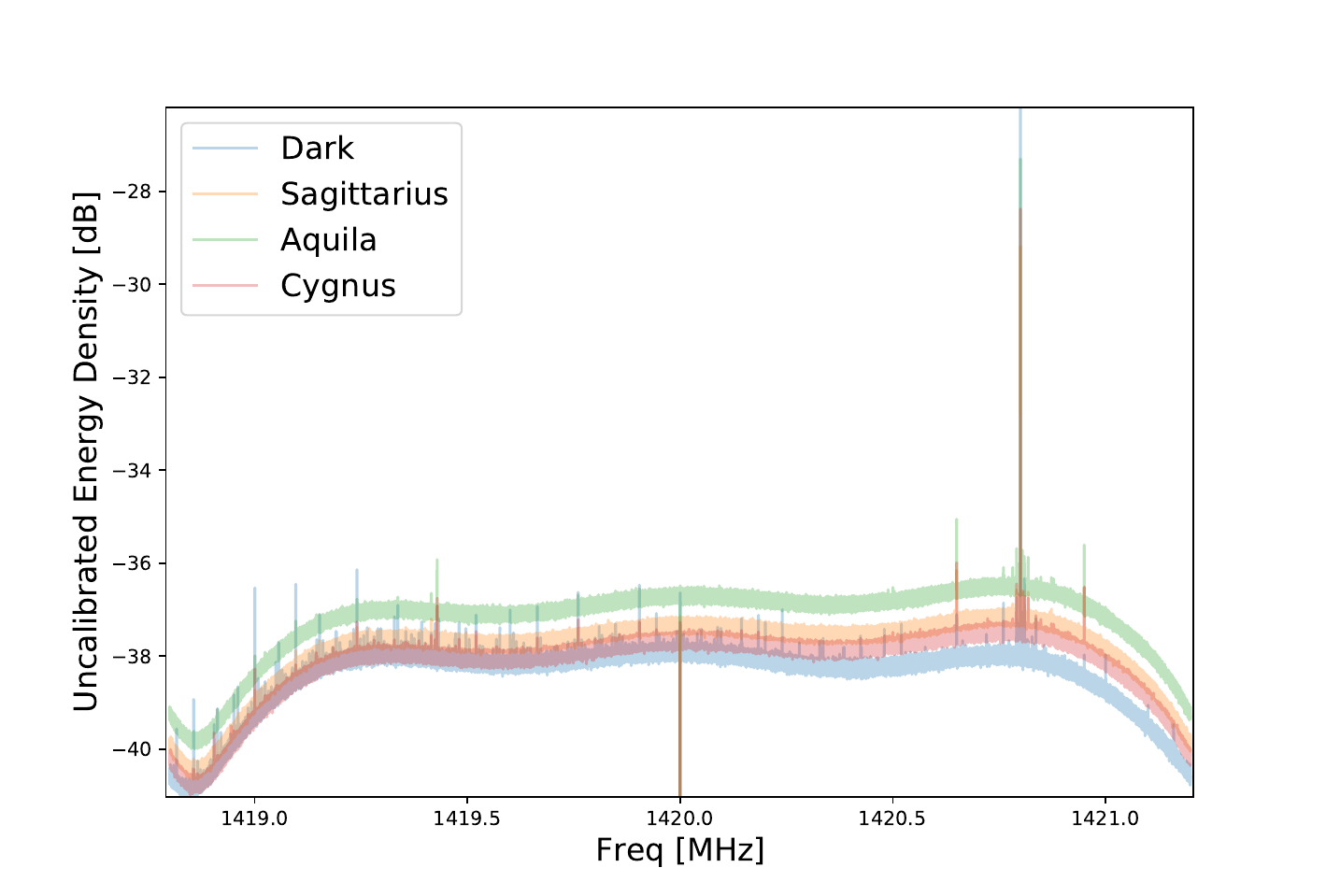}
	    \caption{The uncalibrated spectra we measured. These spectra are obtained by stacking all the frames together, without any correction to the difference in exposure time. As shown in Table.~\ref{tab: pointing-specification}, the Aquila is measured for the longest time, and thus the gives the highest integrated power (energy). We note that the y-axis is normalized arbitrarily, so that only the relative power between different frequencies can be compared. }
	    \label{fig: raw-spectra}
	\end{figure}
    The observation began at 21 July, 2023. 
    After transporting the system from the laboratory to the site, and the on site test runs to ensure system stability,
    the observing run started at $20:18$ (local time, GMT+8), and eventually ended at $21:13$.
    The run started and ended with $10$ minutes dark frame each. 
    The description of dark frames would be detailed below in Section.~\ref{sect: dark-frame}.

    The sky was slightly cloudy that night. 
    The weather was partially influenced by the aftermath of the tropical cyclone Talim that approached Hong Kong, closest at $16-17$ July, 2023. 
    We have splitted the $10$ minutes observation into two (almost) consecutive $\sim 5$ minutes exposures.
    A photo showing the configuration of our setup during observation is available in Fig.~\ref{fig: wok-in-onbservation}.
    The raw spectra we obtained are shown in Fig.~\ref{fig: raw-spectra}.
    We note that the power in the baseband center ($f = 1420.0$ MHz) is manually suppressed to avoid numerical artifacts.
    Furthermore, saturated strong line-alike emission is clearly visible at $f=1420.8$ MHz.
    The line noise persists among different pointings, suggesting that the line noise is difficult to remove. 
    
    \section{Data Analysis} \label{sect: data-pipeline}
    \subsection{Calibration}
    \subsubsection{Dark Frame and Frequency Response of the System} \label{sect: dark-frame}
    We take $2$ dark frames, each with an exposure time of $10$ minutes, before the start of the experiment and the end of experiment. 
    The dark frames are taken as the sky dark, where each exposure is pointed towards sky position at least $30^\circ$ away from the Milky way. 
    The pointing of the dark frames in the $2$ exposures are different, but with the same intention of avoiding pickup of other terrestrial radio source either by the main beam and the side lobes of the antenna. 
    With this consideration, the dark frames are taken at (RA, DEC) = $(210^\circ, 19^\circ)$ (galactic coordinate $(\ell = 10^\circ, b =70^\circ)$) and (RA, DEC) = $(330^\circ,10^\circ)$ (galactic coordinate $(\ell = 70^\circ, b =-30^\circ)$). 

    We find a consistent line emission at $f=1420.8$ MHz that saturated our system, so that the secondary peaks of this line emission are also visible. 
    \footnote{For very intense emission that exceed the dynamical range of the $8$ bit I/Q voltage samples, all the time steps that exceed the range maximum are represented by the same $8$ bit number (255), which effectively form a top-hat function in the time domain. The Fourier transform of the top-hat function is the sinc function, which is featured by multiple symmetric peaks with decaying amplitude around the main peak. }
    As the emission persists across all pointings, we infer that the emission is from nearby electrical equipments in the buildings of the university.
    We conclude that such noise is unavoidable, and we decide the mask out the data at a window of size $\Delta f_{\rm mask} = \pm 20$ kHz around the location of the main peak.

    We also identify that the frequency-integrated total power measured by the system fluctuates at $\lesssim 1\%$ level across all the $4$ seconds frames during the measurement. 
    To mitigate the stability problem, we extract only the relative power across different frequency from each frame. 
    This is done by dividing the magnitude power spectrum $P(f)$ by the frequency-integrated total power, so that:
    \begin{equation}
        \Tilde{P}(f) \equiv P(f) / \sum_f P(f).
    \end{equation}
    
    \subsubsection{Polynomial Baseline Removal}
        \begin{figure}
	    \centering
	    \includegraphics[width=\textwidth]{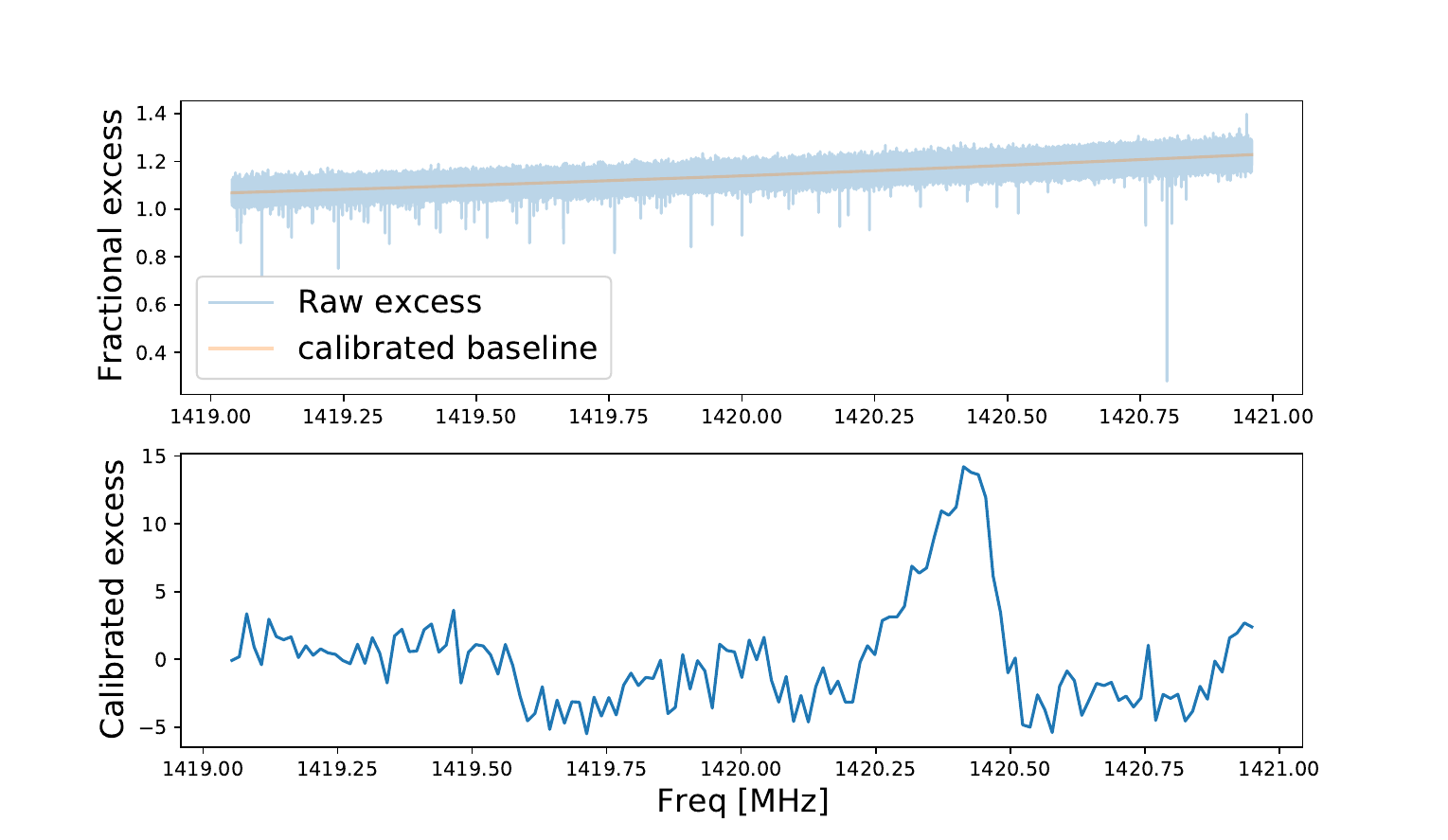}
	    \caption{The spectrum of fractional power excess of the Sagittarius exposure after correcting for the dark. On the top panel, we show the fractional excess before the removal of a baseline, whereby a residual spectral slope is clearly identified. On the bottom panel, we show the fractional power excess after removing a quadratic baseline and re-binning with frequency bins of size $\Delta f=13.7$ kHz. An unambiguous peaky feature around $f = f_{\rm 21} = 1420.4$ MHz is clearly visible. } 
	    \label{fig: calibration}
	\end{figure}
    After identifying the noise properties using the dark frames, we analysis the pointings towards the Milky Way. 
    As explained, the power spectrum are normalized by the total power to mitigate the system stability problem. 
    After normalization, the normalized power spectrum of the pointings are then divided by the normalized power spectrum of the dark frames, to obtain the fractional power excess in the Milky Way pointings:
    \begin{equation} \label{eq: fractional-power-excess}
        \mathcal{S}(f) \equiv \frac{\Tilde{P}_{\rm MW}(f)}{\Tilde{P}_{\rm dark}(f)}
    \end{equation}
    This quantity is plotted in the top panel in Fig.~\ref{fig: calibration}. 
    As can be seen, there remains a persistent spectral slope.
    We assume that the fractional power excess is contributed by two parts:
    \begin{equation} \label{eq: power-excess-model}
        \mathcal{S}(f) = \mathcal{S}_{\rm sys}(f) + \mathcal{S}_{\rm 21}(f)
    \end{equation}
    The systematics $\mathcal{S}_{\rm sys}(f)$ can be modelled by a second order polynomial with an almost vanishing second order derivative, while the $21$ cm emission $\mathcal{S}_{\rm 21}(f)$ has a peaky feature around $f_{21} = 1420.4$ MHz (subject to redshift)
    Therefore, we mask the frequency range $f_{21} \pm 0.2 $ MHz to fit a second order polynomial.
    After determining the best fit polynomial, we subtract the measured fractional power excess by the best fit polynomial to obtain the $21$ cm emission using Eq.~\ref{eq: power-excess-model}.
    The final results is then binned by $13.7$ kHz bins so as to average out the remaining white noise.
    We use the Sagittarius measurement to demonstrate this calibration process, and the results is plotted in Fig.~\ref{fig: calibration}.
    On the top panel of Fig.~\ref{fig: calibration}, we show the fractional excess spectrum (defined via Eq.~\ref{eq: fractional-power-excess}). 
    The slope in the fractional excess spectrum shows the nonlinear response of the system, up to $20\%$ deviation from the dark. 
    This deviation is subsequently captured by a quadratic polynomial and is removed on the bottom panel.

    We repeat this procedure with all the Milky way pointings, and obtain $3$ Milky way spectra that measure $21$ cm emission along directions on the galactic plane.
    The calibrated spectra are shown in Fig.~\ref{fig: obs-result}.
    
    \subsection{Astronomical Interpretation}
    \subsubsection{Recession Velocity}
    \begin{table}
    \centering
    \begin{tabular}{c|c|c}
            Constellation & galactic $\ell$ [$^\circ$] & Recession Velocity [km s$^{-1}$]\\
            \hline \hline
            Sagittarius & 0  & $-2.8^{+20.3}_{-11.6}$ \\
            \hline
            Aquila  & 40 & $-11.5^{+5.8}_{-8.7}$ \\
            \hline
             Cygnus & 80 & $-23.1^{+11.6}_{-17.4}$ \\
             \hline
        \end{tabular}
        \caption{The recession velocity measured in each of the pointings. The error bars in recession velocities are defined by the full width half maximum. Note that each of the error bars is a composition of systematics error and measurement error, as we do not take into account of the intrinsic velocity dispersion of neutral hydrogen inside the spiral arm during signal modelling. }
        \label{tab: recession-velocity}
    \end{table}
    \begin{figure}
	    \centering
	    \includegraphics[width=\textwidth]{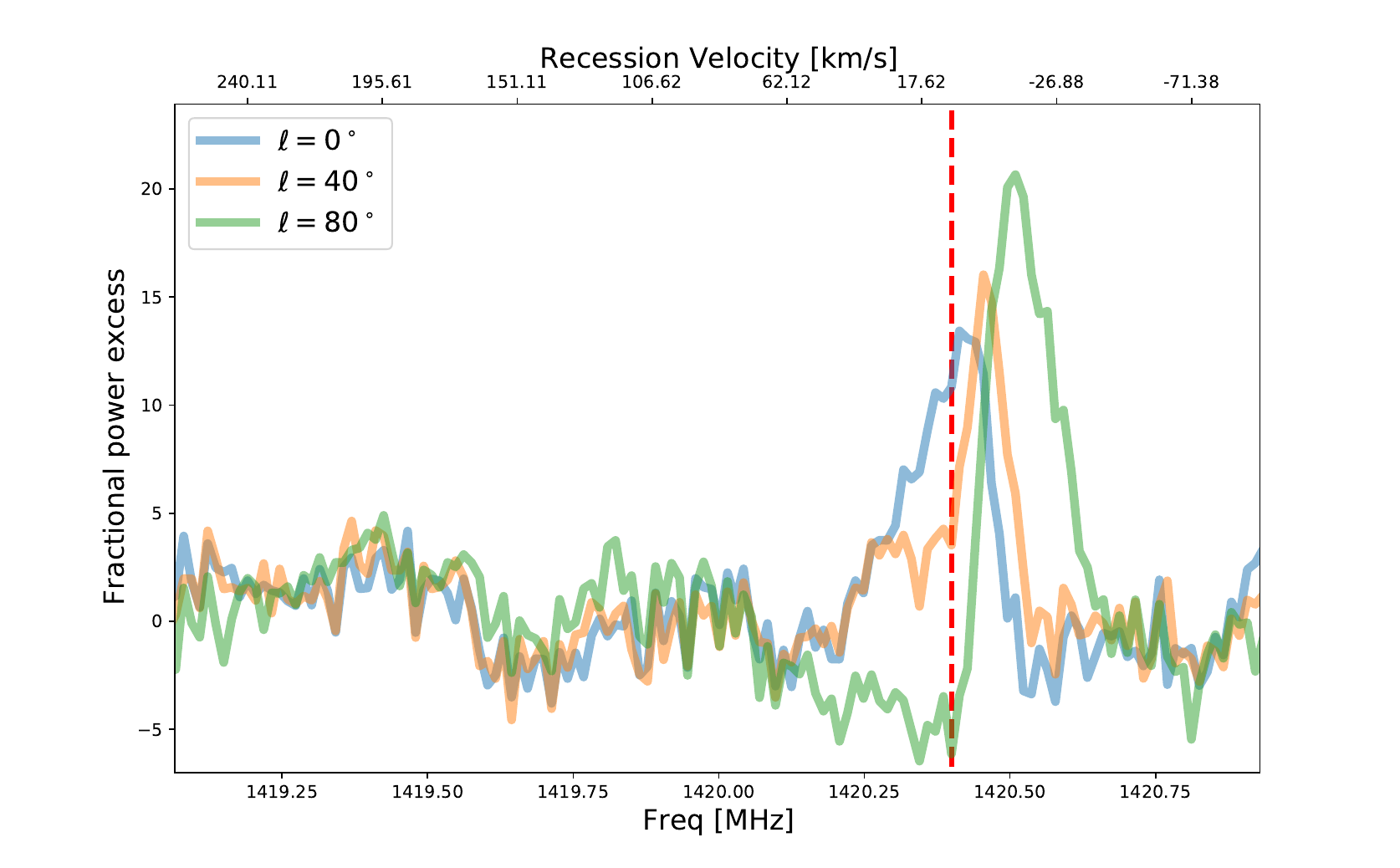}
	    \caption{The calibrated fractional power excess of all the exposures. The vertical red dashed line marks the rest frame frequency of the target $21$ cm emission. One can clearly see that the maximal blueshift of $\Delta v \approx -20$ km s$^{-1}$ along the direction of Cygnus ($\ell \approx 80^{\circ}$). Note the fractional power excess on the y-axis is indeed proportional to the signal-to-noise ratio (SNR), as it quantifies how much stronger the signal than is the noise floor. Note that the pointing specification in the legend are all subject to manual pointing error of roughly $\pm 5^\circ$ and the telescope beam width of $\pm 12^\circ$. } 
	    \label{fig: obs-result}
	\end{figure}
    Using the above method to extract the redshifted emission peak of the $21$ cm emission, we can derive the corresponding recession velocity of the neutral hydrogen with respect to the Earth. 
    This is as simple as using the Doppler formula:
    \begin{equation}
        \Delta v_{//} = c f_{\rm 21} (\frac{1}{f_{\rm obs}} - \frac{1}{f_{\rm 21}}),
    \end{equation}
    where $c$ is the speed of light. 
    
    However, we must address the complication that the emission line has its intrinsic width. Instead of modeling the shape of the emission line and carefully disentangling the broadening effect from the instrument response and intrinsic width, we choose to attribute the broadening effect as the uncertainty in determining the peak location. We measure the full width half maximum (FWHM) of the 21 cm emission and report the inferred recession velocities of each pointing in Table~\ref{tab: recession-velocity}.

    One should exercise caution when interpreting the error bars associated with the reported values. We acknowledge that part of the error bars are systematic as we use a line (equivalently a Dirac-Delta) to model the emission. Furthermore, a particular pointing can pass through multiple spiral arms of the Milky Way, and the neutral hydrogen along each arm has a different intrinsic velocity. This effect can be observed particularly in the Sagittarius pointing, where the width of the emission is substantially larger than the remaining pointings. In fact, upon comparison with other independent measurements in the same region, multiple unresolved spiral arms along that direction are evident, resulting in the intrinsic width being much smaller than the error bar we report here.
    
    The aforementioned degeneracies among the broadening effects can be resolved with a more sophisticated error model. However, considering the quality of the data, the degeneracies in the model parameters could not be resolved, and Bayesian marginalization would be required to derive useful constraints on recession velocity. In this work, we have adopted the simplest method as outlined above.

    \subsubsection{Discussion: Galactic Rotation Curve} \label{sect: rotation-curve}
    It is tempting to derive the galactic rotation curve using recession velocity data. There is a hobbyist project that claims to have measured the rotation curve using the tangent point method \cite{physicsOpenLab-project,PhysicsOpenLab-rotation-curve}. However, as we will argue below, the precision of the WTH experiment is still insufficient for deriving the rotation curve of the Milky Way (MW). In the following, we will discuss the tangent point method and the technical requirements for measuring the rotation curve.

    To translate recession velocity into the galactic rotation curve, additional information related to the geometry of the MW is needed. 
    These details are necessary because the measured recession velocity only represents the projection of the rotational velocity along the line of sight. 
    Without the additional input about the geometry, which cannot be directly measured by our WTH system, it is impossible to derive any constraints on the rotation curve. The tangent point method, however, is a clever technique that bypasses the need for this extra information about MW's geometry.
    
    The principle behind the tangent point method is that the measured recession velocity reaches its maximum when the direction of the instantaneous orbital velocity aligns perfectly with the line of sight. Assuming a circular orbit for the neutral hydrogen, this corresponds to the tangential velocity of the hydrogen in that orbit. By employing this approach, the degeneracy between velocity components along different directions is removed. 
    The tangent point method has been extensively studied and applied since the last century (see, for example, \cite{tangent-point-method-review,rotation-curve-review,tangent-point-method-systematics,tangent-point-method-systematics-2}). Implementing this method requires the ability to determine the maximum velocity along a specific pointing direction.
    
    As the corresponding circular orbit of the tangent point is usually near the MW center with a smaller orbit radius, the signal-to-noise ratio (SNR) of the line-of-sight signal is instead dominated by the spiral arm that is the closest to Earth, but not the tangent point. 
    Therefore, the SNR attainable for measuring the tangent point velocity is generally lower than that for measuring the main emission peak.
    
    In both Fig.~\ref{fig: obs-result} and Table~\ref{tab: recession-velocity}, we can unambiguously resolve the main peak of the $21$ cm emission, which is contributed by the nearest spiral arm. However, the remaining peaks, including the one corresponding to the tangent point, cannot be measured due to their poor SNR. Therefore, the tangent point method is not applicable to our data.
    
    To obtain the rotation curve in future experiments, several steps can be taken. Firstly, the noise floor needs to be further lowered by conducting the experiment in a more radio-quiet environment. This should be possible in most locations except Hong Kong. Another approach, independent of geographical constraints, is to increase the exposure time to accumulate a higher SNR. 
    In our current setup, the SNR for the peak emission is around $5$ with exposure times of less than $10$ minutes. 
    To resolve the higher-order emission peaks, which are typically an order of magnitude dimmer than the main peak, we would need roughly $10$ times higher SNR. 
    This translates to $100$ times longer exposure time, equivalent to a week of total exposure time. This is a rough estimate, as more sophisticated approaches, such as modeling the emission spectra, can also be employed to improve the SNR besides using longer exposures. 
    Template-fitting-based methods like matched-filtering can be adopted to capture the higher-order peaks would likely increase the detectability as well.
    
    \section{Conclusion and Future Improvements} \label{sect: conclusion}
    In this paper, we outline and demonstrate the possibility of constructing a simple radio telescope using kitchenware, primarily for educational and outreach purposes. We show that with a low-cost setup of $\$150$, it is possible to observe the galactic neutral hydrogen and the associated Doppler shift even in noisy urban environments. 
    However, it is important to implement suitable noise mitigation measures.

    We highlight the design considerations involved in configuring the WTH setup, including the choice of reflector geometry, the coupling of the dipole antenna, and the signal pathing. 
    To make the project more accessible to the general public, we choose to use commercial electronic parts instead of tailor-made components. 
    We also explain a few noise mitigation measures that can be generally applicable to many similar projects involving the detection of faint radio waves.
    
    In addition to hardware, we provide detailed information about the data processing pipeline, which includes data compression and dark frame subtraction procedures. This information would be useful for anyone interested in reproducing this system.
    
    We are currently reviewing and planning for future upgrades to this WTH system. This includes the development of a field gateway programmable array (FPGA) readout system for efficient real-time data compression and pre-processing.
    We are also exploring the possibility of extending the system into an intensity interferometer. 
    This would require careful treatment to handle the relative phase delay on the two radio antenna system, which necessitates the careful synchronization of clock systems. 
    The WTH design would serve as a baseline for the future, upon which more sophisticated modifications and experiments can be conducted.

    \section*{Acknowledgement}
    The authors acknowledge that the costs involved in this project are fully covered by themselves without external funding. We would also like to thank Prof. George F. Smoot and Prof. Kam-Biu Luk for their useful comments and discussion, and for approving the use of the laboratory facilities in the Hong Kong University of Science and Technology.
    
    \appendix
    \section{Determination of the Geometry of the Wok} \label{app: wok-geometry}
    The curvature of the wok could be easily characterized if it resembles a parametric curve.
    In this work, we consider both the spherical model and the parabolic model.
    This two models can be distinguished by measuring the maximal perimeter of the cross section of the wok $S_{\rm max}$. 
    
    For the quadratic equation of the form $h = ax^2 + bx + c$, the focal point is located at center of the parabola $x_f=-b/2a$, at a height of:
    \begin{equation} \label{eq: focal-parabola}
        h_f = \frac{4ac-b^2 + 1}{4a}
    \end{equation}
    To ensure the wok is indeed a parabola, we first measure the radius $r$, the depth $d$ and the maximal perimeter of the cross sections $S_{\rm max}$ of the wok.
    For a wok of radius $r$ and a depth $d$, the parabola parameters can be determined by:
    \begin{align}
        h(x= \pm r) = d &; \, h(x) > 0 \, \forall x \\
        \implies a = d/r^2 &; \,\, b=c=0 
    \end{align}
    If the wok indeed have a parabolic shape, the maximal perimeter $S_{\rm max}$ of the wok would be:
    \begin{equation}
    \begin{split}
        S_{\rm max} &= 2\int^{x=r}_{x=-r} d( \sqrt{x^2+h(x)^2} ) = 4\int^{r}_{0} \frac{x + h(x)\frac{dh}{dx} }{\sqrt{x^2+h(x)^2}} \, dx \\
        &= 4\sqrt{d^2+r^2}.
    \end{split} 
    \end{equation}
    For $r=30.5$ cm and $d$ = $17.8$ cm, $S_{\rm max} \approx 141 $ cm.
    
    Although the measured value for $S_{\rm max}$ lies around $144$ cm, which is close to $141$ cm, the practical measurement of the \textbf{maximum} perimeter is hard to achieve. $\mathcal{O}(1)$ centimeter of measurement error is expected. 
    Therefore, we also consider the spherical model to see if the corresponding $S_{\rm max}$ is significantly different from the parabolic model.
    
    For the spherical model, the radius $R$ of a sphere that inscribe a sector of radius $r$ and depth $d$ is:
    \begin{equation}
        R = \frac{d^2 + r^2}{2d}.
    \end{equation}
    The maximal perimeter with a depth $d$ can be calculated with elementary geometry:
    \begin{equation}
        S_{\rm max} = 4R \tan^{-1}\left( \frac{r}{R-d} \right),
    \end{equation}
    which gives $S_{\rm max} = 148$ cm.
    This number is still close to the previous number we got from a parabolic model. 
    Owing to possible deformation and defects during manufacturing process, the wok is unlikely to be either perfectly spherical or parabolic.
    We conclude that it is difficult to accurately describe the geometry of the wok.
    However, both the spherical and parabolic are good approximation to model the wok. 
    For considerations as detailed in Section~\ref{sect: gain-simulation}, we would approximate the wok with a sphere throughout the work for mathematical convenience.
    
	\bibliographystyle{plain}
	\bibliography{ref,ref-wavedm}
\end{document}